\documentclass[12pt,a4paper]{article}
\usepackage{amsbsy}
\usepackage{amsmath}
\usepackage{graphicx}

\title{Fermion Zero Modes for Abelian BPS Monopoles}

\author{
B. Cheng\thanks{bobby.cheng11@imperial.ac.uk}~~~ and~~~ C. Ford\thanks{c.ford@imperial.ac.uk
}
\\
Department of Mathematics\\
 Imperial College London\\
London SW7 2AZ}

\begin{document}\maketitle

\begin{abstract}

Fermion zero modes for abelian BPS monopoles are considered.
In the spherically symmetric case the normalisable
zero modes are determined for
arbitrary
monopole 
charge $N$.
If $N>1$
the zero modes are zero along $N-1$ half-lines emanating from the monopole.

\end{abstract}

\bigskip\noindent
{Keywords: abelian gauge theory, BPS monopoles, Weyl equation, fermion zero modes.}

\bigskip

Fermion zero modes for BPS monopoles can be constructed via the same Nahm transform
used to obtain the monopoles \cite{Nahm:1979yw}.
The construction is, however, cumbersome for magnetic charge $N>2$.
In this letter we obtain zero modes for abelian BPS monopoles.
Our approach is to directly integrate the Weyl equations in three-dimensional space
rather than use  Nahm's method (which has been adapted to
abelian monopoles  in
\cite{Nahm:1980jy}).

The abelian BPS equations read
\begin{equation}\label{bps}
{\bf B}=\nabla\Phi,\end{equation}
where $\Phi$ is a real Higgs field and ${\bf B}$ is a magnetic field
derived from a vector potential ${\bf A}$.
The Maxwell equation
$\nabla\cdot {\bf B}=0$ implies that
the Higgs field $\Phi$ obeys the Laplace equation.
The Higgs field
\begin{equation}\label{higgs}\Phi=\frac{g}{2 \pi}\left(a-\frac{1}{2}
\sum_{i=1}^N{\frac{1}{|{\bf r}-{\bf r}_i|}}\right),\end{equation}
with $a$ and $g$ constant,
is harmonic away from $N$ singularities ${\bf r}_i$ ($i=1,2,...,N$).
Physically, the system comprises $N$ Dirac monopoles
each with magnetic charge $g$ 
interacting with a Higgs field. Here $a$ fixes the asymptotic value
of the Higgs field.
Consider the Weyl operators 
\begin{equation}D=eI_2\Phi+i {\boldsymbol \sigma} \cdot(-i\nabla+e{\bf A})~~~~~~
D^{\dagger}= eI_2\Phi-i{\boldsymbol \sigma}\cdot(-i\nabla+e{\bf A})\end{equation}
where $e$ is the electric charge of the fermion
and $\boldsymbol\sigma=(\sigma_1,\sigma_2,\sigma_3)$.
These Weyl operators assume a real Yukawa coupling in Minkowski space.
However, identifying $\Phi$ as $A_0$ they are also Weyl operators for self-dual
monopoles defined in Euclidean space.

The Dirac quantisation condition
requires
$eg=2\pi n$ with $n$ integer.
If 
$$eg=2\pi$$ and $a>0$,
$D^\dagger$ has $N$ normalisable zero modes.
In the $N=1$ case we have
\begin{equation}
\Phi={g\over{2\pi}}\left(a-{1\over{2 r}}\right),~~~~
{\bf A}=
{g\over{4\pi}}{y{\bf e}_x-x{\bf e}_y\over{r(r-z)}}=-{g(1+\cos\theta){\bf e}_{\phi}\over{4\pi r\sin\theta}},
\end{equation}
taking the origin as the location of the monopole
and $(r,\theta,\phi)$ denote spherical polar coordinates.
Here
 the Dirac string lies  on the positive $z$-axis.
  
 One can verify that
 \begin{equation}
 \psi={e^{-ar}\over{\sqrt r}}\left(
 \begin{array}{l}-\sin{(\theta/2)}\\
 \cos(\theta/2)e^{i\phi}\end{array}
 \right)
 \end{equation}
 is a zero mode of $D^\dagger$
 (the components of $D^\dagger$ in spherical polar coordinates are
  given in the appendix). $D$ has no zero modes.
This result can be obtained by taking the large $r$ limit of the Jackiw-Rebbi zero
mode \cite{Jackiw:1975fn,GonzalezArroyo:1995ce} for the basic $SU(2)$ BPS
monopole
after performing a gauge transformation which diagonalises
the Higgs field.
The fermion
density $\psi^\dagger\psi=e^{-2ar}/r$ is spherically symmetric
and has an integrable singularity at the monopole centre.

The general $N$ case is more complicated. However, in the spherically symmetric
case where the positions of the $N$ monopoles coincide the Higgs field is 
\begin{equation}
\label{HiggsN}
\Phi={g\over{2\pi}}\left(a-{N\over{2r}}\right)\end{equation}
and  ${\bf A}$ is the $N=1$ potential multiplied by $N$.
Here $D^\dagger$ has $N$ normalisable zero modes:
\begin{equation}
\psi^{m}=r^{{1\over2}(N-2)}e^{-ar}\left(
 \begin{array}{l}\!\!-\sin^{N-m+1}{(\theta/2)}\cos^{m-1}(\theta/2)e^{i(m-1)\phi}\!\!\\~~~~~
 \sin^{N-m}(\theta/2)
 \cos^m(\theta/2) e^{im\phi}\end{array}
 \right)~~~m=1,2,...,N.\end{equation}
 These solutions resemble (in particular with respect to the $\theta$ and $\phi$ dependence)
 known solutions of the Dirac equation in the background of
 abelian monopoles \cite{Kazama:1976fm}.
 However, our solutions are written directly in terms of trigonometric
 functions rather than spherical harmonics\footnote{
 If $N$ is even the zero modes can be expressed in terms of standard spherical
  harmonics.
 If $N$ is odd spin-weighted or monopole harmonics \cite{Wu:1976ge}
 are required.}.
 Our solutions are normalisable 
 with 
  $L^2$ norm
 $4\pi (N-m)!(m-1)!(2a)^{-(N+1)}$.
 As the zero modes
 $\psi^m$ are annihilated by the Hamiltonian
 it is not clear to us whether the presence of the Higgs field cures
 the self-adjointness problem \cite{ 
 Kazama:1976fm,Lipkin:1969ck} 
 associated with monopole Hamiltonians.
 To address this issue one needs to study the scattering states \cite{Kazama:1976fm}.

Note that the densities $\psi^m{}^\dagger\psi^m$ are not spherically
symmetric for $N>1$.
For $N=2$ the first zero mode is zero along the positive $z$-axis
while the second mode is zero on the negative $z$-axis.
By taking a suitable
linear combination of $\psi^1$ and $\psi^2$
one can obtain
a zero mode with a zero along any half-line emanating from the monopole;
the zero mode is axially symmetric about the axis on which the half-line lies.
The  $N>1$ zero modes are zero along $N-1$ half-lines. Our $\psi^1$
and $\psi^N$ have zeros of strength $N-1$ on the positive and negative
$z$-axes, respectively; the remaining $N-2$ modes have lower strength zeros
on both the positive and negative $z$ axes.
Again one can adjust the directions of the $N-1$ half lines by 
taking different linear combinations of the $N$ zero modes.
For a discussion of zeros of fermion zero modes in a different context see
\cite{Bruckmann:2004dk}.

In general, the $N>2$ zero modes are not axially symmetric
even though the $\psi^m$ are all axially symmetric about the $z$-axis.
If $\Psi$ is a linear  combination of the $\psi^m$, the density
$\Psi^\dagger\Psi$ 
 has the form
\begin{equation}\Psi^\dagger\Psi=r^{N-2}e^{-2ar}f(\theta,\phi),
\end{equation}
where $f$ is a function of $\theta$ and $\phi$.
For $N>2$ one can see that the zero modes vanish at the position of the monopole
and the zero modes 
peak somewhere on the sphere $r=(N-2)/2a$.
The function $f(\theta,\phi)$ has up to $N-1$ zeros;
if there are fewer than $N-1$ zeros these are repeated zeros associated with coincident half-lines.
Examination of $f(\theta,\phi)$
for several zero modes indicates that $f(\theta,\phi)$ has a single peak.
For example, the $N=3$ zero mode $\Psi=\psi^1+\psi^2+\psi^3$ yields an $f(\theta,\phi)$
function
with two zeros and a  maximum on the equator $\theta={\pi/2}$ (these three points
are equally spaced on the equator).
The maximum of $f(\theta,\phi)$ is a point on the unit sphere
except for some axially symmetric solutions where $f(\theta,\phi)$
peaks on a circle.
In Figure 1  spherical plots of $f(\theta,\phi)$ are given for an $N=2$ and an $N=3$ zero mode.

\begin{figure}
\includegraphics[scale=0.3]{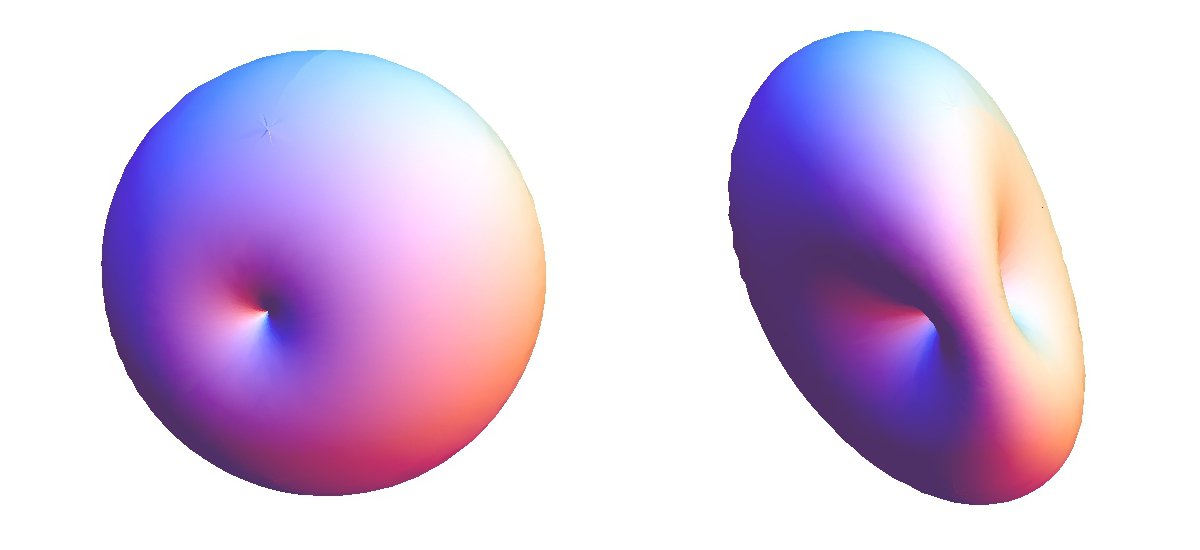}
\caption{
Spherical plots of $f(\theta,\phi)$ for an $N=2$ and $N=3$ zero mode (the scale depends
on the normalisation and hence $a$). 
The left plot shows $f(\theta,\phi)$  for an $N=2$ zero mode ($\Psi=\psi^1+\psi^2$) which has one zero and is axially symmetric.
Two zeros are visible for the $N=3$ zero mode ($\Psi=\psi^1+\psi^2-i\psi^3$) on the right.
Both zero modes have a single peak.
}
\end{figure}

van Baal \cite{VanBaal:2002rt}
has given an ansatz
which provides a   solution of the Weyl equation
for any solution of the abelian BPS equation:
\begin{equation}
\label{vanbaal1}       
\psi_{vB}=
D\left(
\begin{array}{l}
w\\0\end{array}
\right)
\end{equation}
satisfies $D^\dagger\psi_{vB}=0$ where
\begin{equation}
\label{vanbaal2}
\Phi={g\over{2\pi}}{\partial\over{\partial z}}\log w,~~~~~
{\bf A}=-{g\over{2\pi}}{\bf k}\times\nabla \log w 
\end{equation}
and
$\log w$ satisfies the Laplace equation.
This works as the BPS equation implies $D^\dagger D$ is a scalar
and (\ref{vanbaal2}) gives
$D^\dagger D w=0$.
Taking

\noindent
$\log w=
{1\over2}N\log(r-z)+az$
yields our $\Phi$ and ${\bf A}$.
However, $\psi_{vB}$ is not normalisable\footnote{
Remarkably,
(\ref{vanbaal1}) does provide one normalisable
zero mode for a different class of Higgs fields;
here $\Phi$ has $2N$ singularities representing $N$ positively charged
and $N$ negatively charged monopoles \cite{VanBaal:2002rt,Bruckmann:2003ag}. }
though for $a=0$ it agrees with our zero mode $\psi^1$.
Indeed, $\psi^1$ is the $a=0$ van Baal solution multiplied by $e^{-ar}$.
As the van Baal construction does not rely on spherical symmetry this
approach may provide information about the
$a=0$ limit of the general case where the $N$ monopoles are separated.

We have considered fermion zero modes for BPS monopoles and have obtained solutions for arbitrary magnetic charge $N$.
The spherically symmetric abelian case we have solved may provide a model for non-abelian
magnetic bags;
although higher charge $SU(2)$ BPS monopoles are never spherically symmetric,
solutions with approximate spherical symmetry may emerge for large $N$
\cite{Bolognesi:2005rk}.
It would be interesting to investigate the extent to which our zero modes approximate
the fermion zero modes of magnetic bags.

 \appendix
 
 \section*{Appendix}
The components of $D^\dagger$ associated with (\ref{HiggsN})
are ($eg=2\pi$)
\begin{flalign*}
(D^\dagger)_{11}&=a - \frac{N}{2r} - \cos\theta\,\frac{\partial}{\partial r} + \frac{\sin\theta}{r}\,\frac{\partial}{\partial \theta}
\\
(D^\dagger)_{12}&=e^{-i\phi}
\left[
- \sin\theta\,\frac{\partial}{\partial r} - \frac{\cos\theta}{r}\,\frac{\partial}{\partial \theta} + \frac{i}{r\sin\theta}\,\frac{\partial}{\partial \phi} + \frac{N}{2}\frac{(1+\cos\theta)}{r\sin\theta}
\right]\\
(D^\dagger)_{21}&=e^{i\phi}
\left[
- \sin\theta\,\frac{\partial}{\partial r} - \frac{\cos\theta}{r}\,\frac{\partial}{\partial \theta} - \frac{i}{r\sin\theta}\,\frac{\partial}{\partial \phi} - \frac{N}{2}\frac{(1+\cos\theta)}{r\sin\theta}
\right]\\
(D^\dagger)_{22}&=a - \frac{N}{2r} + \cos\theta\,\frac{\partial}{\partial r}- \frac{\sin\theta}{r}\,\frac{\partial}{\partial \theta}.
\end{flalign*}

\end{document}